\documentclass[aps,pra,twocolumn,showpacs,bibnotes]{revtex4-1}

\usepackage{graphicx}

\graphicspath{{pict/}{}}

\newcommand{\be}{\begin{equation}}
\newcommand{\ee}{\end{equation}}

\newcommand\pictc[5]{\vspace*{-0mm}\begin{figure}
                       \centerline{
                       \includegraphics[width=#1\columnwidth,height=0.7\textheight,keepaspectratio]{#3}}
                       \protect\caption{\protect\label{fig:#4} #5}\vspace{-0mm}
                    \end{figure}            }

\newcommand\pict[4][0.97]{\pictc{#1}{h}{#2}{#3}{#4}}
\newcommand\rpict[1]{\ref{fig:#1}}

\newcommand\leqt[1]{\protect\label{eq:#1}}
\newcommand\reqtn[1]{\ref{eq:#1}}
\newcommand\reqt[1]{(\reqtn{#1})}

\newcommand\lsect[1]{\protect\label{sect:#1}}
\newcommand\rsect[1]{\ref{sect:#1}}

\newcounter{Fig}

\begin{document}

\title{Nonlinear suppression of time-reversals in ${\cal PT}$-symmetric optical couplers}

\author{Andrey A. Sukhorukov}
\author{Zhiyong Xu}
\author{Yuri S. Kivshar}

\affiliation{Nonlinear Physics Centre, Research School of Physics and Engineering,
Australian National University, Canberra ACT 0200, Australia}

\begin{abstract}
We reveal a generic connection between the effect of time-reversals and nonlinear wave dynamics in systems with parity-time (${\cal PT}$) symmetry, considering a symmetric optical coupler with balanced gain and loss where these effects can be readily observed experimentally.
We show that for intensities below a threshold level, the amplitudes oscillate between the waveguides, and the effects of gain and loss are exactly compensated after each period due to {periodic time-reversals}. For intensities above a threshold level, nonlinearity suppresses periodic time-reversals leading to the symmetry breaking and a sharp beam switching to the waveguide with gain. Another nontrivial consequence of linear ${\cal PT}$-symmetry is that the threshold intensity remains the same when the input intensities at waveguides with loss and gain are exchanged.
\end{abstract}


\maketitle

\section{Introduction} \lsect{intro}

The study of different physical systems exhibiting parity-time (${\cal PT}$) symmetry has attracted a lot of attention in the past few years.  The original motivation stems from the initial ideas of the generalization of quantum mechanics suggested by Bender {\em et al.}~\cite{Bender:1998-5243:PRL, Bender:2002-270401:PRL, Bender:2003-1095:AMJP, Bender:2007-947:RPP}, who demonstrated that a wide class of Hamiltonians, even though non-Hermitian, can still exhibit entirely real eigenvalue spectra provided that they obey the so-called parity-time requirements or ${\cal PT}$ symmetry. A necessary (but not sufficient) condition for a Hamiltonian to be ${\cal PT}$-symmetric is that its potential $V(x)$ satisfies the condition $V(x) = V^*(-x)$.
It was suggested that in optics the refractive index modulation combined with gain and loss regions can play a role of
complex ${\cal PT}$-symmetric
potentials~\cite{Ruschhaupt:2005-L171:JPA, El-Ganainy:2007-2632:OL, Klaiman:2008-80402:PRL}.
It should be noted that the beam dynamics in directional couplers composed of waveguides with gain and loss was originally described theoretically already two decades ago in Ref.~\cite{Chen:1992-239:IQE}, however the recently identified analogy with ${\cal PT}$ symmetry property has stimulating
extensive theoretical~\cite{Berry:2008-244007:JPA, Makris:2008-103904:PRL, Longhi:2009-123601:PRL,
Bendix:2009-30402:PRL, West:2010-54102:PRL, Longhi:2010-22102:PRA, Ramezani:1005.5189:ARXIV} and experimental~\cite{Guo:2009-93902:PRL, Ruter:2010-192:NPHYS} studies.

As was demonstrated in the original study of directional couplers with gain and loss~\cite{Chen:1992-239:IQE}, such structures can offer benefits for all-optical switching in the nonlinear regime, lowering the switching power and attaining sharper switching transition. Recently, these conclusions were complimented by the prediction of unidirectional switching and exact analytical solution describing the switching dynamics in nonlinear ${\cal PT}$ symmetric couplers~\cite{Ramezani:1005.5189:ARXIV}.

In this paper, we show that the origin of nonlinear switching in the ${\cal PT}$-symmetric directional couplers is related to the effect of suppression of time-reversals. This conclusion is based on the symmetry analysis which is applicable to a broad class of nonlinear local responses, including in particular the cases of cubic (as considered in Refs.~\cite{Chen:1992-239:IQE, Ramezani:1005.5189:ARXIV}) or saturable responses. This is important in view of possible experimental realizations of such couplers in different material systems with various nonlinear response characteristics. For example, linear ${\cal PT}$-symmetric couplers have been demonstrated based on LiNbO$_{3}$ platform~\cite{Ruter:2010-192:NPHYS}, and this material possesses photorefractive nonlinearity with saturable response.

We show that although nonlinearity always breaks the ${\cal PT}$-symmetry conditions for asymmetric wave profiles even at arbitrarily small intensity levels, the effects of gain and loss are exactly compensated and ${\cal PT}$-symmetric dynamics is preserved {\em on average} due to periodic time-reversals, for intensities below a certain threshold. In contrast, for intensities above a threshold, nonlinear self-action suppresses time-reversals and ${\cal PT}$-symmetric dynamics is broken both locally and globally, resulting in the asymmetric wave localization in the region with gain.

The paper is organized as follows. In Sec.~\rsect{model} we discuss our model of a nonlinear ${\cal PT}$-symmetric coupler.  Stationary solutions of this model are found and analyzed in Sec.~\rsect{stationary}, whereas Sec.~\rsect{symmetry} provides the symmetry analysis underpinning general conclusions on the types of nonlinear dynamics on the coupler. Finally, results of numerical simulations which confirm and complement the analytical analysis are presented Sec.~\rsect{numerics}, and Sec.~\rsect{conclusions} concludes the paper.

\section{Model equations} \lsect{model}

We describe the propagation of waves in a ${\cal PT}$-symmetric optical coupler by the equations for the mode amplitudes at the first and second waveguides. We use a set of coupled-mode equations which include additional terms accounting for Kerr-type nonlinearity~\cite{Chen:1992-239:IQE, Ruter:2010-192:NPHYS}:
\begin{equation} \leqt{DNLS}
   \begin{array}{l} {\displaystyle
       i \frac{d a_1}{d z} + i \rho a_1 + C a_2 + G(|a_1|^2) a_1 = 0, \quad
   } \\*[9pt] {\displaystyle
       i \frac{d a_2}{d z} - i \rho a_2 + C a_1 + G(|a_2|^2) a_2 = 0,
   } \end{array}
\end{equation}
where
$z$ is the propagation distance, $a_1$ and $a_2$ are the mode amplitudes, $\rho = \rho_1 = - \rho_2$ defines the rates of loss in the first waveguide and gain in the second waveguide, $C$ is the coupling coefficient between the modes of two waveguides, and function $G$ characterizes the nonlinear response.
We assume with no loss of generality that $C>0$, since for negative $C$ it is possible to make the transformation $a_2 \rightarrow -a_2$ and $C \rightarrow -C$.
We note that a recently analyzed model for Bose-Hubbard dimer in the mean-field approximation considered in Ref.~\cite{Graefe:1003.3355v1:ARXIV} may appear to be similar to the optical case, however the mean-field nonlinear terms are different which response does not depend on the total amplitude scaling in contrast to Kerr-type nonlinearity in Eqs.~\reqt{DNLS}.

In the following, we consider the values of gain/loss coefficient below the linear ${\cal PT}$-symmetry breaking threshold~\cite{Ruter:2010-192:NPHYS}, $\rho < C$.
In order to analyze nonlinear dynamics, it
 is convenient to represent the mode amplitudes in the following form,
\begin{equation} \leqt{a12}
   \begin{array}{l} {\displaystyle
       a_{1}=\sqrt{I(z)} \cos[\theta(z)] \exp[+i \varphi(z)/2] \exp[i \beta(z)],
   } \\*[9pt] {\displaystyle
       a_{2}=\sqrt{I(z)} \sin[\theta(z)] \exp[-i \varphi(z)/2] \exp[i \beta(z)],
   } \end{array}
\end{equation}
where $I$ is the total intensity,
$\theta$ and $\varphi$ define the relative intensities and phases between the two input waveguides, and $\beta$ is the overall phase.
After substituting Eq.~\reqt{a12} into Eq.~\reqt{DNLS}, we derive the closed system of evolution equations for $I$, $\theta$, and $\varphi$:
\begin{equation} \leqt{dI}
   \begin{array}{l} {\displaystyle
      \frac{d I}{d z} = - 2 \rho I \cos(2 \theta) , \quad
      \frac{d \theta}{d z} = \rho \sin(2 \theta) - C \sin\varphi ,
   } \\*[9pt] {\displaystyle
      \frac{d \varphi}{d z} = G( I \cos^2\theta ) - G( I \sin^2\theta ) - 2 C \cot(2 \theta) \cos\varphi ,
   } \end{array}
\end{equation}
and additional equation for $\beta$:
\begin{equation} \leqt{beta}
 \frac{d \beta}{d z} = \frac{1}{2}\left[G( I \cos^2\theta )/2 + G( I \sin^2\theta ) \right]
 + \frac{C \cos\varphi}{\sin(2 \theta)}.
\end{equation}

\pict{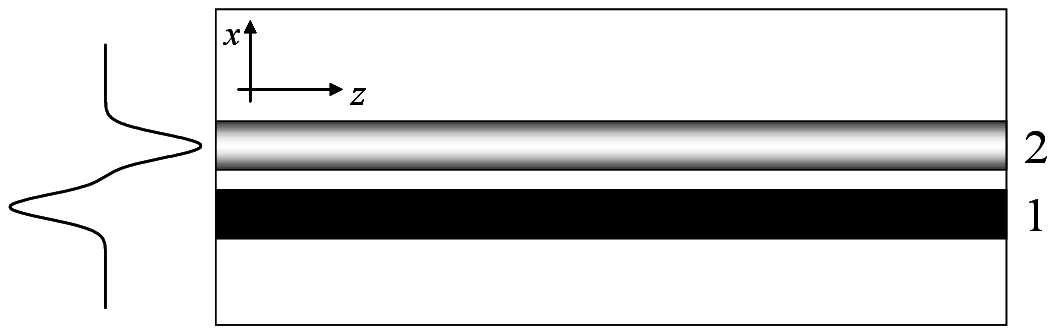}{scheme}{
Scheme of nonlinear ${\cal PT}$-symmetric directional coupler with balanced loss in waveguide 1 and gain in waveguide 2.
The arrow indicates the propagation direction ($z$).
}

\section{Stationary solutions} \lsect{stationary}

Before performing the analysis of strongly nonstationary dynamics, we first analyze stationary solutions of Eqs.~\reqt{dI}. We find that such nonlinear modes correspond to the phase space points $(I=I_0,\theta = \pi/4,\varphi=\varphi_\pm)$ and $\beta = \beta_\pm z$, where $\beta_\pm = G(I_0 / 2) + C \cos(\varphi_\pm)$, $\sin(\varphi_\pm) = \rho / C$, and $\cos(\varphi_\pm) = \mp [ 1 - (\rho/C)^2 ]^{1/2}$.
We perform the linear stability analysis by considering weakly perturbed stationary solutions in the form:
\begin{equation} \leqt{aPerturb}
   \begin{array}{l} {\displaystyle
         a_{j} = \exp(i \beta_\pm z) \{ \sqrt{I_0/2} \exp[ i (-1)^{j+1} \varphi_\pm/2 ]
   } \\*[9pt] {\displaystyle\quad
     + u_{j} \exp(i p z) + v_{j}^\ast \exp(-i p^\ast z)\} ,
   } \end{array}
\end{equation}
where $u_j$ and $v_j$ are the perturbation amplitudes and $p$ defines the growth rate.
We identify non-trivial perturbation modes with eigenvalues
\begin{equation} \leqt{p}
    p = \pm 2 C \left[\cos^2(\varphi) - \tilde{\gamma} \cos(\varphi)\right]^{1/2},
\end{equation}
where $\tilde{\gamma} = G'(I_0 / 2) I_0 / ( 2 C)$ and prime stands for the derivative. The fixed point in phase space is a stable center, ${\rm Im}(p) = 0$, if (i)~$\tilde{\gamma} < \tilde{\gamma}_{\rm cr}$ and $\varphi = \varphi_-$ or (ii)~$\tilde{\gamma} > - \tilde{\gamma}_{\rm cr}$ and $\varphi = \varphi_+$, where $\tilde{\gamma}_{\rm cr} = |\cos(\varphi_\pm)|$. If these conditions are not satisfied, then the fixed point is a saddle, and a mode amplitude corresponding to ${\rm Im}(p) < 0$ grows exponentially indicating the onset of instability.

\section{Symmetry properties and dynamical scenarios} \lsect{symmetry}

In order to describe the features of nonstationary dynamics, we
identify an important symmetry property of the model equations.
After performing the complex conjugation of Eq.~\reqt{DNLS} and comparing it to the original equations, we conclude that for any solution $a_j(z)$,
\begin{equation} \leqt{PTa}
        \widetilde{a}_1(z_{0+}) = a_2^\ast(z_{0-}) e^{i \delta}, \,
        \widetilde{a}_2(z_{0+}) = a_1^\ast(z_{0-}) e^{i \delta}
\end{equation}
is also a solution of Eq.~\reqt{DNLS} for arbitrary constants $z_0$ and $\delta$, where $z_{0\pm} = z_0 \pm z$.
This transformation represents the action of ${\cal PT}$ operator, where parity operator (${\cal P}$) corresponds to exchange of waveguide numbers and time operator (${\cal T}$) defines the reversal of propagation direction.
Using notations of Eq.~\reqt{a12}, we express the transformation in Eq.~\reqt{PTa} as
\begin{equation} \leqt{PT}
   \begin{array}{l} {\displaystyle
        \widetilde{I}(z_{0+}) = I(z_{0-}),\quad
        \widetilde{\varphi}(z_{0+}) = \varphi(z_{0-}),
   } \\*[9pt] {\displaystyle
        \widetilde{\theta}(z_{0+}) = \pi/2 - \theta(z_{0-}),\quad
        \widetilde{\beta}(z_{0+}) = \delta - \beta(z_{0-}).
   } \end{array}
\end{equation}
We notice that if $z_0 = z_m$ where
\begin{equation} \leqt{zm}
   \theta(z_m) = \pi/4,
\end{equation}
then we can choose the free parameter as $\delta = 2 \beta(z_0)$, and solution transforms into itself at $z=z_m$.
This happens because the intensity distribution is symmetric, $|a_1(z_m)|^2 \equiv |a_2(z_m)|^2$, and accordingly nonlinearity does not break the ${\cal PT}$-symmetry condition at $z=z_m$.
Since the original and transformed solutions satisfy the same evolution equation, it follows that
\begin{equation} \leqt{PTm}
   \begin{array}{l} {\displaystyle
        {I}(z_{m+}) = I(z_{m-}), \,
        {\theta}(z_{m+}) = \pi/2 - \theta(z_{m-}), \,
   } \\*[9pt] {\displaystyle
        {\varphi}(z_{m+}) = \varphi(z_{m-}), \,
        {\beta}(z_{m+}) = 2 \beta(z_m) - \beta(z_{m-}),
   } \end{array}
\end{equation}
for $z_{m\pm} = z_m \pm z$ and any $z_m$ which satisfies condition in Eq.~\reqt{zm}.
According to Eq.~\reqt{PTm}, the dynamics starting from $z_m$ in positive ($+z$) and negative ($-z$) directions is exactly equivalent, subject to the effective exchange of waveguide numbers [Eq.~\reqt{PTa}], and this is a nontrivial consequence of linear ${\cal PT}$ symmetry in the nonlinear regime.
It also follows from Eqs.~\reqt{dI} and~\reqt{zm} that
\begin{equation} \leqt{dIzm}
  \left.\frac{d I}{d z}\right|_{z=z_m} = 0 .
\end{equation}
The physical interpretation of this important result is that {\em the system exhibits effective time-reversal when the total intensity reaches the maximum or minimum values}, where time-reversal (${\cal T}$) corresponds to change of the propagation direction ($z$).
We use this result to reveal the general properties of nonstationary wave dynamics.
 %for arbitrarily long propagation distances ($-\infty < z < +\infty$),

First, we show that for any conditions at $z=0$ and considering $-\infty< z <+\infty$, the dependence $I(z)$ should contain a minimum (or a stationary point), where $d I / d z = 0$ and $d^2 I / d z^2 \ge 0$ at $z_{\rm min}$. This can be proven as a contradiction: (i)~Depending on the sign of $d I / d z$ at $z=0$, we choose the positive or negative propagation direction corresponding to decreasing intensity (if such direction does not exist, $z=0$ is already a minimum). (ii)~As the intensity is decreasing, the effect of nonlinearity is diminishing (we assume the regular nonlinear response with $G(I) \rightarrow 0$ as $I \rightarrow 0$),
and
 %If a minimum intensity value is not reached,
the system approaches linear propagation regime. (iii)~However, it was shown that in the linear regime the intensity dependence always contains a minimum when loss coefficient is below the linear ${\cal PT}$ breaking threshold~\cite{El-Ganainy:2007-2632:OL, Guo:2009-93902:PRL, Ruter:2010-192:NPHYS}. Since there always exists $z_{\rm min}$, it follows from Eq.~\reqt{PTm} that all trajectories in phase space are symmetric with respect to transformation $\theta \rightarrow \pi/2 - \theta$.
We note that for special initial conditions, at stage (ii) the solution may be asymptotically approaching a saddle type fixed point (this is not possible for the only other type of stationary solutions in our model corresponding to a stable center). As we have shown above, fixed points only appear when $\theta=\pi/4$, exactly fulfilling the condition of Eq.~\reqt{zm}. Therefore, such case formally corresponds to a limit $z_{\rm min} \rightarrow +\infty$. However, such separatrix trajectories approaching a saddle point are inherently unstable and this special case is practically unreachable and can be neglected.

Second, we reveal that for arbitrary nonlinear response functions, all solutions belong to two classes: (i)~periodic solutions, where the intensities and relative phases in two waveguides are exactly restored after each period ($z\rightarrow z+z_p$), or (ii)~solutions where the total intensity grows without bound due to nonlinearly-induced symmetry breaking.
Let us prove that solutions are periodic if intensity is bounded, i.e. when there exists $z_{\rm max}$ where $d I / d z = 0$ and $d^2 I / d z^2 < 0$ (we again neglect the special case of a separatrix trajectory approaching a saddle point). Then, according to relation in Eq.~\reqt{dIzm}, Eq.~\reqt{PTm} should be satisfied simultaneously for $z_m = z_{\rm max}$ and $z_m = z_{\rm min}$, i.e.
\begin{equation} \leqt{Imm}
        {I}(z_{\rm min} - z) = I(z_{\rm min} + z), \,
        {I}(z_{\rm max} - z) = I(z_{\rm max} + z)
\end{equation}
Making a variable transformation $z \rightarrow (z_{\rm min} - z)$ and $z \rightarrow (z_{\rm max}-z)$ in the first and second relations in Eq.~\reqt{Imm} respectively, we obtain
\begin{equation} \leqt{Imm2}
        {I}(z) = I(2 z_{\rm min} - z), \,
        {I}(z) = I(2 z_{\rm max} - z) .
\end{equation}
Applying the second relation in Eq.~\reqt{Imm2} recursively after the first one, we find that
\begin{equation} \leqt{Imm3}
        {I}(z) = I(2 z_{\rm max} - 2 z_{\rm min} + z) .
\end{equation}
This means that the solution is periodic, with the period equal to $z_p = 2 | z_{\rm max} - z_{\rm min}|$.
We can determine the location of extrema points ($z_{\rm max}$ and $z_{\rm min}$) on the phase plane. It follows from Eq.~\reqt{dI} and~\reqt{zm} that maxima ($z_{\rm max}$) correspond to $\theta = \pi/4$ and $\varphi_- < \varphi < \varphi_+$, and minima ($z_{\rm min}$) to $\theta = \pi/4$ and $\varphi < \varphi_-$ or $\varphi > \varphi_+$.

\pict{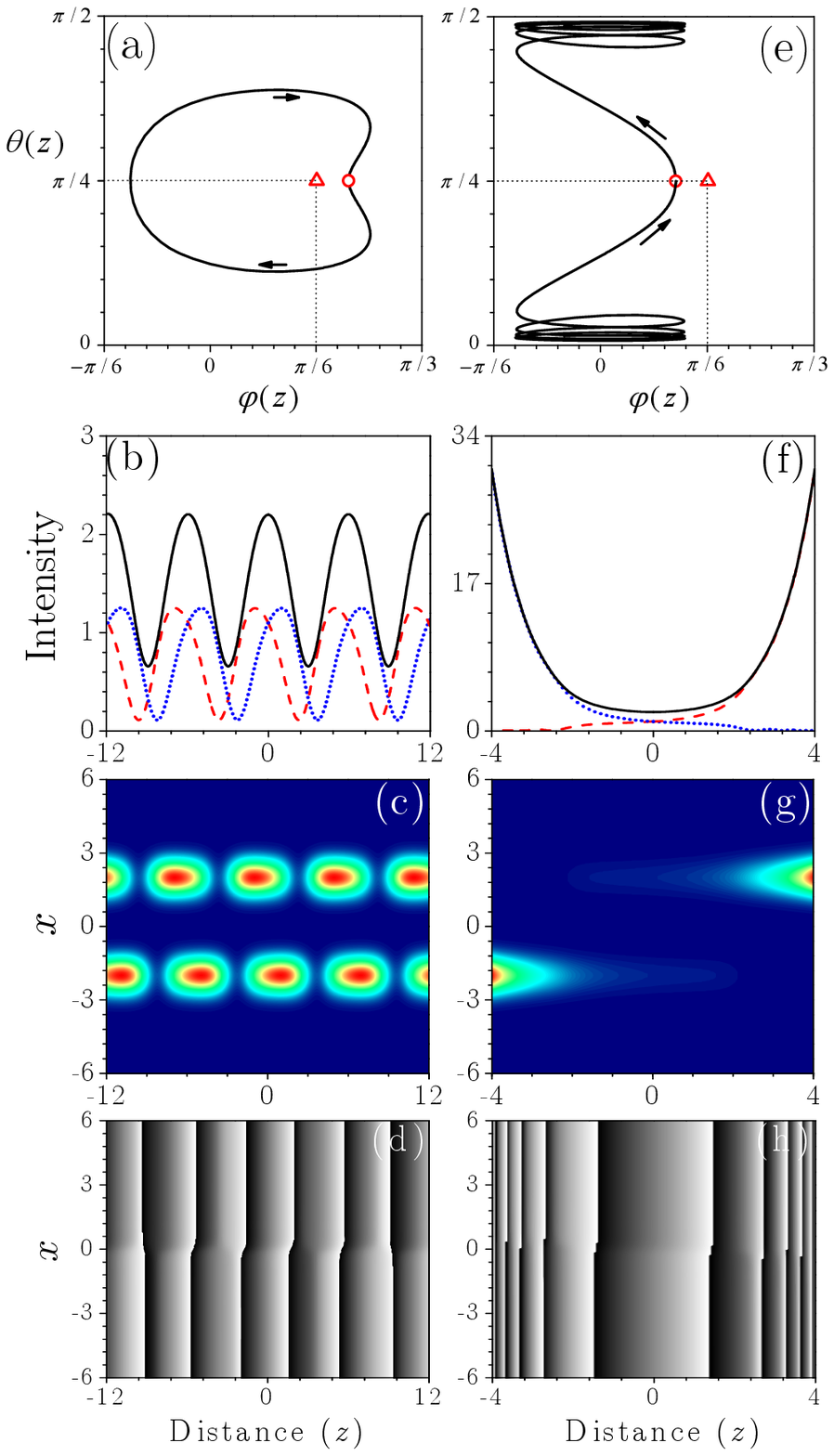}{dynamics}{(Color online)
System dynamics for different initial conditions: (a)-(d)~$\varphi=\pi/6-\pi/20$ and (e)-(h)~$\varphi=\pi/6+\pi/20$.
(a),(e)~Trajectories in the phase plane ($\theta$, $\varphi$). Red open circle marks the point at $z=0$, and open triangle marks
the unstable stationary solution with $\varphi_-=\pi/6$.
(b),(f)~Intensity dependencies on propagation distance in the first (dotted line) and second (dashed) waveguides, solid line show the sum of individual intensities.
(c),(g) and (d),(h) show the intensity and phase evolution along the propagation direction.
For all the plots, $\rho=0.5$ and $I(z=0)=2.2$.}

Based on these general predictions, we can reveal a remarkable property. The type of nonlinear dynamics (periodic or unbounded) remains the same if we swap the intensities between the two waveguides [see Eq.~\reqt{PTm}]. In particular, we can couple light at the input just to the first waveguide with loss, or to the second waveguide with gain, and the type of dynamics would be the same. This is a counter-intuitive result, since in the first case the total intensity will initially decrease, whereas in the second case the total intensity will be growing. However, in both cases the type of dynamics will be determined only by the initial intensity level. This is a highly nontrivial consequence of linear ${\cal PT}$-symmetry in the strongly nonlinear regime.

\pict{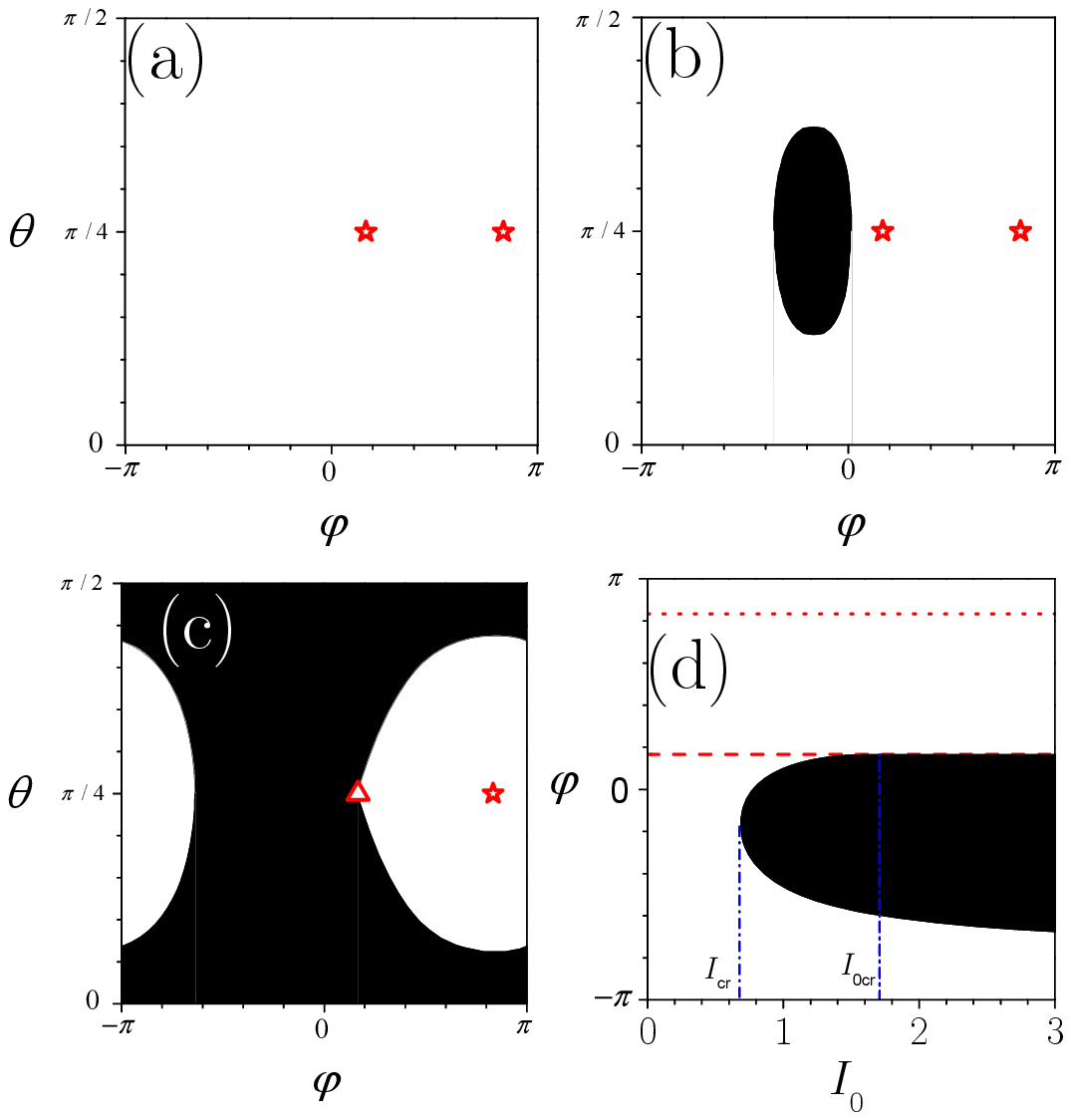}{PTcouplerRegions}{
(Color online) Regions of \emph{PT}-symmetry (white shading) and symmetry breaking with
nonlinear switching (black shading) in the plane of initial
conditions:
(a)-(c) the plane ($\theta$, $\varphi$) for different input intensities:
(a)~$I=0.1$, (b)~$I=0.8$, and (c)~$I=2.2$. Stars and triangles mark stable and unstable stationary solutions, respectively.
(d) the plane ($I$, $\varphi$) for $\theta = \pi/4$.
Dashed and dotted lines mark the stationary points at $\varphi_\pm$.
For all the plots $\rho=0.5$.}

\section{Numerical results for Kerr nonlinearity} \lsect{numerics}

We now complement the general analytical results with numerical examples. To be specific, we consider the Kerr-type nonlinear response function 
\begin{equation} \leqt{G}
  G(I) = \gamma I, 
\end{equation}
where $\gamma > 0$ for self-focusing nonlinearity. Then, by introducing the transformation $z \rightarrow z C$ and $a_j \rightarrow a_j \sqrt{C/\gamma}$, we can scale the values of coefficients to unity, $C=1$ and $\gamma = 1$, and we use these values in numerical simulations. It follows from analytical analysis presented in Sec.~\rsect{stationary} that the stationary point at $\varphi_+$ is always stable, whereas at $\varphi_-$ the instability appears for $I_0 > I_{\rm 0cr} = 2 [ 1 - \rho^2 ]^{1/2}$. We present in Fig.~\rpict{dynamics} two examples of system dynamics when the initial condition (at $z=0$) is chosen in the vicinity of the unstable point, $\varphi_- = \pi/6$ for $\rho = 0.5$ and $I_0 = 2.2 > I_{\rm 0cr} = \sqrt{3}$.
In both examples we choose $\theta = \pi/4$, such that $z_m = 0$ according to Eq.~\reqt{zm}. In the first example, we set the value of $\varphi(z=0)$ slightly larger than $\varphi_-$, and therefore according to the general analytical results the initial condition corresponds to a maximum ($z_{\rm max} = 0$), and the solution should be periodic. Indeed, this is confirmed by numerical simulations in Fig.~\rpict{dynamics}(a)-(d). We see that the trajectory in phase space rapidly moves away from the unstable stationary point, but then returns to the initial location after a full period. Completely different dynamics is observed in the second example [Fig.~\rpict{dynamics}(e)-(h)], where we set the value of $\varphi(z=0)$ slightly less than $\varphi_-$. In this case, the initial condition corresponds to a minimum ($z_{\rm min} = 0$), and the solution does not have to be periodic. Indeed, we see that the total intensity grows without bound and light becomes concentrated in a single waveguide at $|z| \rightarrow \infty$. These examples illustrate two types of the system dynamics.

Next, we perform comprehensive numerical studies to determine the dynamical regimes for arbitrary initial conditions, which may not correspond to the vicinity of stationary points. In Fig.~\rpict{PTcouplerRegions} we show the regions of the initial conditions corresponding to periodic (white shading) and unbounded (black shading) solutions, for $\rho = 0.5$. The plots in Figs.~\rpict{PTcouplerRegions}(a-c) are shown in the plane ($\theta$, $\varphi$) for different input intensities, and Fig.~\rpict{PTcouplerRegions}(d) --- in the plane ($I$, $\varphi$) for $\theta = \pi/4$.
For the input intensity below a critical threshold ($I < I_{\rm cr}$), all initial conditions correspond to periodic solutions preserving ${\cal PT}$-symmetry on average. Note that the threshold value is lower than the instability threshold for the stationary mode, i.e. $I_{\rm cr} < I_{\rm 0cr}$. As the input intensity exceeds the threshold, there appears a range of initial conditions corresponding to nonlinear suppression of ${\cal PT}$ symmetry reversals. On the other hand, according to the analytical results, the region between the dashed and dotted lines [corresponding to the stationary points with $\varphi_\pm$) in Fig.~\rpict{PTcouplerRegions}(d)] will always remain white, since these initial conditions correspond to periodic solutions at arbitrary intensities.

\pict{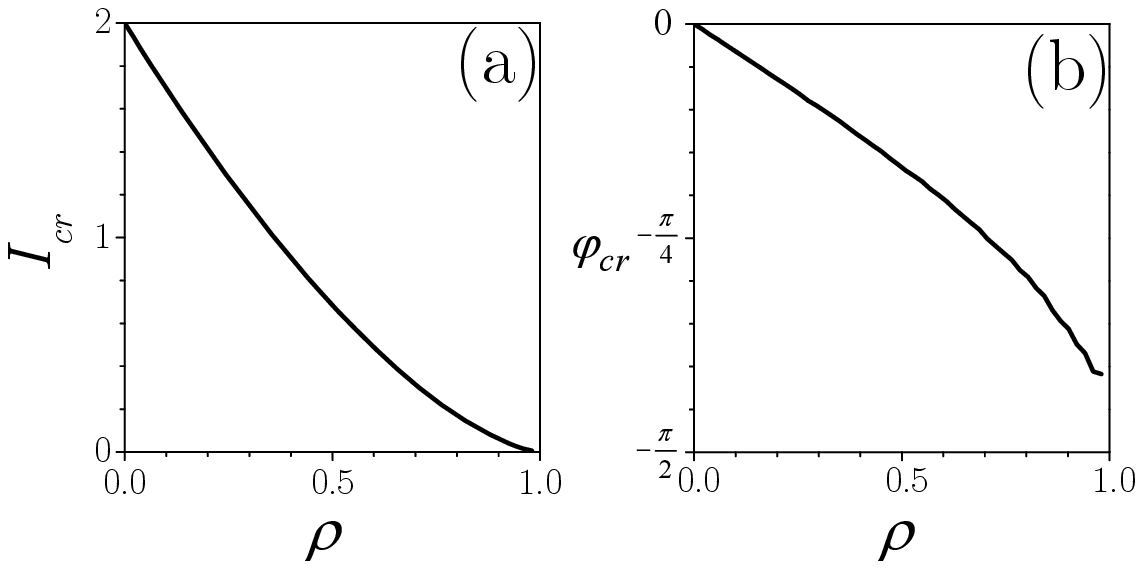}{critical}{
(a)~Dependence of the minimal critical intensity required for nonlinear switching vs. the gain coefficient $\rho$.
(b)~The corresponding value of the input phase $\varphi_{\rm cr}$ for the nonlinear switching.
}

Finally, in Fig.~\rpict{critical} we present the dependence of the minimal input intensity ($I_{\rm cr}$) on gain coefficient and the corresponding phase difference ($\varphi_{\rm cr}$) required for nonlinear switching. According to the analytical results, the minimal intensity should always correspond to the condition $\theta_{\rm cr} = \pi/4$, and our numerical results confirm this conclusion. As follows from Fig.~\rpict{critical}(a),
the threshold for nonlinear switching is drastically reduced for larger gain/loss coefficients, which is crucial for the practical applications of
such couplers in all-optical signal processing.

\section{Conclusions} \lsect{conclusions}

We have described analytically and demonstrated numerically the effect of nonlinearity-induced ${\cal PT}$-symmetry breaking in directional waveguide couplers with balanced gain and loss. We have revealed that time-reversals can support average balance between gain and loss despite local ${\cal PT}$-symmetry breaking, whereas suppression of time-reversals at stronger nonlinearities results in switching and light concentration in a region with gain.
Our results may offer a new insight and suggest different possibilities for all-optical control of the beam switching and amplification in nonlinear photonic structures containing loss and gain elements~\cite{Chen:1992-239:IQE, Musslimani:2008-30402:PRL, Musslimani:2008-244019:JPA, Dmitriev:2010-2976:OL, Ramezani:1005.5189:ARXIV}.

This work has been supported by the Australian Research Council through Discovery projects.


\begin{thebibliography}{10}

\bibitem{Bender:1998-5243:PRL}
C.~M. Bender and S. Boettcher, Phys. Rev. Lett. {\bf 80}, 5243 (1998).

\bibitem{Bender:2002-270401:PRL}
C.~M. Bender, D.~C. Brody, and H.~F. Jones, Phys. Rev. Lett. {\bf 89}, 270401
  (2002); Erratum: Phys. Rev. Lett. {\bf 92}, 119902 (2004).

\bibitem{Bender:2003-1095:AMJP}
C.~M. Bender, D.~C. Brody, and H.~F. Jones, Am. J. Phys. {\bf 71}, 1095 (2003).

\bibitem{Bender:2007-947:RPP}
C.~M. Bender, Rep. Prog. Phys. {\bf 70}, 947 (2007).

\bibitem{Ruschhaupt:2005-L171:JPA}
A. Ruschhaupt, F. Delgado, and J.~G. Muga, J. Phys. A {\bf 38}, L171 (2005).

\bibitem{El-Ganainy:2007-2632:OL}
R. {El-Ganainy}, K.~G. Makris, D.~N. Christodoulides, and Z.~H. Musslimani,
  Opt. Lett. {\bf 32}, 2632 (2007).

\bibitem{Klaiman:2008-80402:PRL}
S. Klaiman, U. Guenther, and N. Moiseyev, Phys. Rev. Lett. {\bf 101}, 080402
  (2008).

\bibitem{Chen:1992-239:IQE}
Y.~J. Chen, A.~W. Snyder, and D.~N. Payne, IEEE J. Quantum Electron. {\bf 28},
  239 (1992).

\bibitem{Berry:2008-244007:JPA}
M.~V. Berry, J. Phys. A {\bf 41}, 244007 (2008).

\bibitem{Makris:2008-103904:PRL}
K.~G. Makris, R. {El-Ganainy}, D.~N. Christodoulides, and Z.~H. Musslimani,
  Phys. Rev. Lett. {\bf 100}, 103904 (2008).

\bibitem{Longhi:2009-123601:PRL}
S. Longhi, Phys. Rev. Lett. {\bf 103}, 123601 (2009).

\bibitem{Bendix:2009-30402:PRL}
O. Bendix, R. Fleischmann, T. Kottos, and B. Shapiro, Phys. Rev. Lett. {\bf
  103}, 030402 (2009).

\bibitem{West:2010-54102:PRL}
C.~T. West, T. Kottos, and T. Prosen, Phys. Rev. Lett. {\bf 104}, 054102
  (2010).

\bibitem{Longhi:2010-22102:PRA}
S. Longhi, Phys. Rev. A {\bf 81}, 022102 (2010).

\bibitem{Ramezani:1005.5189:ARXIV}
H. Ramezani, T. Kottos, R. El~Ganainy, and D.~N. Christodoulides, arXiv {\bf
  {\mdseries 1005.5189}} (2010).

\bibitem{Guo:2009-93902:PRL}
A. Guo, G.~J. Salamo, D. Duchesne, R. Morandotti, M. {Volatier-Ravat}, V.
  Aimez, G.~A. Siviloglou, and D.~N. Christodoulides, Phys. Rev. Lett. {\bf
  103}, 093902 (2009).

\bibitem{Ruter:2010-192:NPHYS}
C.~E. Ruter, K.~G. Makris, R. {El-Ganainy}, D.~N. Christodoulides, M. Segev,
  and D. Kip, Nature Physics {\bf 6}, 192 (2010).

\bibitem{Graefe:1003.3355v1:ARXIV}
E.~M. Graefe, H.~J. Korsch, and A.~E. Niederle, arXiv {\bf \mdseries
  1003.3355v1} (2003).

\bibitem{Musslimani:2008-30402:PRL}
Z.~H. Musslimani, K.~G. Makris, R. {El-Ganainy}, and D.~N. Christodoulides,
  Phys. Rev. Lett. {\bf 100}, 030402 (2008).

\bibitem{Musslimani:2008-244019:JPA}
Z.~H. Musslimani, K.~G. Makris, R. {El-Ganainy}, and D.~N. Christodoulides, J.
  Phys. A {\bf 41}, 244019 (2008).

\bibitem{Dmitriev:2010-2976:OL}
S.~V. Dmitriev, A.~A. Sukhorukov, and Yu.~S. Kivshar, Opt. Lett. {\bf 35}, 2976
  (2010).

\end{thebibliography}
\end{document}